\documentclass[prb,aps,showpacs,twocolumn,unsortedaddress]{revtex4}
\usepackage{graphics,bm}
\usepackage{amssymb}
\usepackage{epsfig}
\usepackage{epsf}
\usepackage[usenames]{color}

\begin{document}

\title{Spin pumping by a field-driven domain wall}

\author{R.A. Duine}

\affiliation{Institute for Theoretical Physics, Utrecht
University, Leuvenlaan 4, 3584 CE Utrecht, The Netherlands}
\date{\today}

\begin{abstract}
We present the theory of spin pumping by a field-driven domain
wall for the situation that spin is not fully conserved. We
calculate the pumped current in a metallic ferromagnet to first
order in the time derivative of the magnetization direction.
Irrespective of the microscopic details, the result can be
expressed in terms of the conductivities of the majority and
minority electrons and the dissipative spin transfer torque
parameter $\beta$. The general expression is evaluated for the
specific case of a field-driven domain wall and for that case
depends strongly on the ratio of $\beta$ and the Gilbert damping
constant. These results may provide an experimental method to
determine this ratio, which plays a crucial role for
current-driven domain-wall motion.
\end{abstract}


\maketitle

\def\bx{{\bf x}}
\def\bk{{\bf k}}
\def\half{\frac{1}{2}}
\def\args{(\bx,t)}

\section{Introduction} Adiabatic quantum pumping of electrons in quantum dots
\cite{bruder1994,brouwer1998} has recently been demonstrated
experimentally for both charge \cite{switkes1999} and spin
\cite{watson2003}. Currently, the activity in this field is mostly
concentrated on the effects of interactions \cite{sharma2001},
dissipation \cite{cohen2003}, and non-adiabaticity
\cite{strass2005}. Complementary to these developments, the
emission of spin current by a precessing ferromagnet
--- called spin pumping --- has been studied theoretically and
experimentally in single-domain magnetic nanostructures
\cite{tserkovnyak2002, mizukami2001,costache2006}. One of the
differences between spin pumping in single-domain ferromagnets and
quantum pumping in quantum dots is that in the latter the
hamiltonian of the electronic quasi-particles is manipulated
directly, usually by varying the gate voltage of the dot. In the
case of ferromagnets, however, it is the order parameter --- the
magnetization direction
--- that is driven by an external (magnetic) field. The coupling
between the order parameter and the current-carrying electrons in
turn pumps the spin current \cite{tserkovnyak2005}. The opposite
effect, i.e., the manipulation of magnetization with spin current,
is called spin transfer
\cite{slonczewski1996,berger1996,tsoi1998,myers1999}.

Recently, the possibility of manipulating with current the
position of a magnetic domain wall via spin transfer torques has
attracted a great deal of theoretical
\cite{bazaliy1998,rossier2004,
tatara2004,zhang2004,waintal2004,barnes2005,thiaville2005,rebei2005,ohe2006,xiao2006,tserkovnyak2006,
kohno2006,piechon2006,duine2006,duine2007} and experimental
\cite{grollier2003,tsoi2003,yamaguchi2004,
klaui2005,beach2006,hayashi2007,yamanouchi2004,yamanouchi2006}
interest. Although the subject is still controversial
\cite{tatara2004,barnes2005}, it is by now established that in the
long-wavelength limit the equation of motion for the magnetization
direction $\bm{\Omega}$, which in the absence of current describes
damped precession around the effective field $- \delta E_{\rm MM}
[ \bm{\Omega}]/(\hbar \delta \bm{\Omega})$, is given by
\begin{eqnarray}
\label{eq:LLGwithSTTs} && \left( \frac{\partial }{\partial t} +
{\bf v}_{\rm s} \cdot \nabla  \right) \bm{\Omega} - \bm{ \Omega}
\times
 \left( - \frac{\delta E_{\rm MM} [ \bm{\Omega}]}{\hbar \delta \bm{\Omega}}\right)
 \nonumber \\
 &&=
 -  \bm{\Omega} \times
  \left( \alpha_{\rm G} \frac{\partial }{\partial t}+ \beta {\bf v}_{\rm s} \cdot \nabla \right) \bm{\Omega}~,
\end{eqnarray}
and contains, to lowest order in spatial derivatives of the
magnetization direction, two contributions due the presence of
electric current.

The first is the reactive spin transfer torque
\cite{bazaliy1998,rossier2004}, which corresponds to the term
proportional to $\nabla \bm{\Omega}$ on the left-hand side of the
above equation. It is characterized by the velocity ${\bf v}_{\rm
s}$ that is linear in the curent and related to the external
electric field ${\bf E}$ by
\begin{equation}
 {\bf v}_{\rm s} = \frac{\left( \sigma_\downarrow- \sigma_\uparrow\right) {\bf E}}{|e|\rho_s
 }~,
\end{equation}
where $\sigma_\uparrow$ and $\sigma_\downarrow$ denote the
conductivities of the majority and minority electrons,
respectively, and $\rho_s$ is their density difference. (The
elementary charge is denoted by $|e|$.) The second term in
Eq.~(\ref{eq:LLGwithSTTs}) due to the current is the dissipative
spin transfer torque \cite{footnote1} that is proportional to
$\beta$ \cite{zhang2004,waintal2004,barnes2005}. Both this
parameter, and the Gilbert damping parameter $\alpha_{\rm G}$,
have their microscopic origin in processes in the hamiltonian that
break conservation of spin, such as spin-orbit interactions.

It turns out that the phenomenology of current-driven domain-wall
motion depends crucially on the value of the ratio
$\beta/\alpha_{\rm G}$. For example, for $\beta=0$ the domain wall
is intrinsically pinned \cite{tatara2004}, meaning that there is a
critical current even in the absence of inhomogeneities. For
$\beta/\alpha_{\rm G}=1$ on the other hand, the domain wall moves
with velocity ${\bf v}_{\rm s}$. Although theoretical studies
indicate that generically $\beta \neq \alpha_{\rm G}$
\cite{tserkovnyak2006,kohno2006,piechon2006,duine2007}, it is not
well-understood what the relative importance of spin-dependent
disorder and spin-orbit effects in the bandstructure is, and a
precise theoretical prediction of $\beta/\alpha_{\rm G}$ for a
specific material has not been attempted yet. Moreover, the
determination of the ratio $\beta/\alpha_{\rm G}$ from experiments
on current-driven domain wall motion has turned out to be hard
because of extrinsic pinning of the domain and nonzero-temperature
\cite{yamanouchi2006,duine2006} effects.

In this paper we present the theory of the current pumped by a
field-driven domain wall for the situation that spin is not
conserved. In particular, we show that a field-driven domain wall
in a metallic ferromagnet generates a charge current that depends
strongly on the ratio $\beta/\alpha_{\rm G}$. This charge current
arises from the fact that a time-dependent magnetization generates
a spin current, similar to the spin-pumping mechanism proposed by
Tserkovnyak {\it et al.} \cite{tserkovnyak2002} for nanostructures
containing ferromagnetic elements. Since the symmetry between
majority and minority electrons is by definition broken in a
ferromagnet, this spin current necessarily implies a charge
current. In view of this, we prefer to use the term ``spin
pumping" also for the case that spin is not fully conserved, and
defining the spin current as a conserved current is no longer
possible.

The generation of spin and charge currents by a moving domain wall
via electromotive forces is discussed very recently by Barnes and
Maekawa \cite{barnes2007}. We note here also the work by Ohe {\it
et al.} \cite{ohe2007}, who consider the case of the Rashba model,
and the very recent work by Saslow \cite{saslow2007}, Yang {\it et
al.} \cite{yang2007}, and Tserkovnyak and Mecklenburg
\cite{tserkovnyak2007}. In addition to these recent papers, we
mention the much earlier work by Berger, which discusses the
current induced by a domain wall in terms of an analogue of the
Josephson effect \cite{berger1986}.

Barnes and Maekawa \cite{barnes2007} consider the case that spin
is fully conserved. In this situation it is convenient to perform
a time and position dependent rotation in spin space, such that
the spin quantization axis is locally parallel to the
magnetization direction. As a result of spin conservation, the
hamiltonian in this rotated frame contains now only
time-independent scalar and exchange potential terms. The
kinetic-energy term of the hamiltonian, however, will acquire
additional contributions that have the form of a covariant
derivative. Perturbation theory in these terms then amounts to
performing a gradient expansion in the magnetization direction
\cite{rossier2004}. Hence, the fact that Barnes and Maekawa
consider the case that spin is fully conserved is demonstrated
mathematically by noting that in Eq.~(5) of
Ref.~[\onlinecite{barnes2007}] there are no time-dependent
potential-energy terms. Generalizing this approach to the case of
spin-dependent disorder or spin-orbit coupling turns out to be
difficult. Nevertheless, Kohno and Shibata were able to determine
the Gilbert damping and dissipative spin transfer torques using
the above-mentioned method \cite{kohno2007}. Since Barnes and
Maekawa \cite{barnes2007} consider the situation that spin is
fully conserved, they effectively are dealing with the case that
$\alpha_{\rm G}=\beta=0$. This is because both the Gilbert damping
parameter $\alpha_{\rm G}$ and the dissipative spin transfer
torque parameter $\beta$ arise from processes in the microscopic
hamiltonian that do not conserve spin
\cite{tserkovnyak2006,kohno2006,piechon2006,duine2007}.  Hence,
for the case that $\alpha_{\rm G} = \beta =0$ our results agree
with the results of Barnes and Maekawa \cite{barnes2007}.

The remainder of this paper is organized as follows. In
Sec.~\ref{sec:elecurrent} we derive a general expression for the
electric current induced by a time-dependent magnetization
texture. This general expression is then evaluated in
Sec.~\ref{sec:fddwmotion} for a simple model of field-driven
domain wall motion. We end in Sec.~\ref{sec:disc} with a short
discussion, and present our conclusions and outlook.

\section{Electric Current} \label{sec:elecurrent}
Quite generally, the expectation value of the charge current
density, defined by ${\bf j} = - c \delta H/\delta {\bf A}$ with
$c$ the speed of light, $H$ the hamiltonian, and ${\bf A}$ the
electromagnetic vector potential, is given as a functional
derivative of the effective action
\begin{equation}
\label{eq:expcvaluecurrentfromeffectiveaction}
  \langle {\bf j} (\bx,\tau) \rangle
  = c \frac{\delta S_{\rm eff}}{\delta {\bf A} (\bx,\tau)}~,
\end{equation}
with $\tau$ the imaginary-time variable that runs from $0$ to
$\hbar/(k_{\rm B} T)$. (Planck's constant is denoted by $\hbar$
and $k_{\rm B} T$ is the thermal energy.) First, we assume that
spin is conserved meaning that the hamiltonian is invariant under
rotations in spin space. The part of the effective action for the
magnetization direction that depends on the electromagnetic vector
potential is then given by \cite{rossier2004}
\begin{equation}
\label{eq:effactionadiabatic}
  S_{\rm eff} = \int\!d\tau\!\!\int\!d\bx \left\langle
  j^z_{s,\alpha} (\bx,\tau)  \right\rangle \tilde A_{\alpha'} (\bm{\Omega}
  (\bx,\tau))
  \nabla_\alpha \Omega_\beta (\bx,\tau)~,
\end{equation}
where a summation over Cartesian indices $\alpha,\alpha',\alpha''
\in \{x,y,z\}$ is implied throughout this paper. In this
expression,
\begin{eqnarray}
j_{s,\alpha'}^\alpha (\bx,\tau) &=& \frac{\hbar^2}{4mi} \left[
   \bm{\phi}^\dagger (\bx,\tau) \tau^\alpha \nabla_{\alpha'} \bm{\phi} (\bx,\tau) \right.
   \nonumber \\&& \left.- \left(
   \nabla_{\alpha'} \bm{\phi}^\dagger (\bx,\tau) \right) \tau^\alpha \bm{\phi} (\bx,\tau)
   \right] \nonumber \\ &&
   + \frac{|e| \hbar}{2mc} A_{\alpha'} \bm{\phi}^\dagger (\bx,\tau) \tau^\alpha \bm{\phi} (\bx,\tau)~,
\end{eqnarray} is the spin current, given here in terms of the Grassman coherent state spinor
$\bm{\phi}^\dagger=(\phi^*_\uparrow,\phi^*_\downarrow)$.
Furthermore, $\tau^\alpha$ are the Pauli matrices, and $m$ is the
electron mass. (Note that since we are, for the moment,
considering the situation that spin is conserved there are no
problems regarding the definition of the spin current.) The
expectation value $\langle \cdots \rangle$ is taken with respect
to the current-carrying collinear state of the ferromagnet.
Finally, $\tilde A_{\alpha} (\bm{\Omega})$ is the vector potential
of a magnetic monopole in spin space [not to be confused with the
electromagnetic vector potential ${\bf A} (\bx,\tau)$] that obeys
$\epsilon_{\alpha,\alpha',\alpha''}
\partial \tilde A_{\alpha'}/\partial \Omega_{\alpha''}=\Omega_\alpha$ and is well-known
from the path-integral formulation for spin systems
\cite{auerbachbook}. Eq.~(\ref{eq:effactionadiabatic}) is most
easily understood as arising from the Berry phase picked up by the
spin of the electrons as they drift adiabatically through a
non-collinear magnetization texture
\cite{bazaliy1998,rossier2004}. Variation of this term with
respect to the magnetization direction gives the reactive spin
transfer torque in Eq.~(\ref{eq:LLGwithSTTs}).

The expectation value of the spin current is given by
\begin{equation}
\label{eq:expcvaluecurrentPi}
  \left\langle j_{s,\alpha}^z (\bx,\tau) \right\rangle
  =  \int \!d\tau'\!\!\int\!d\bx' \Pi^z_{\alpha,\alpha'
  } (\bx-\bx';\tau-\tau') \frac{A_{\alpha'}(\bx',\tau')}{\hbar
  c}~.
\end{equation}
The zero-momentum low-frequency part of the response function
$\Pi^z_{\alpha,\alpha'
  } (\bx-\bx';\tau-\tau') \equiv \left\langle
    j^z_{s,\alpha} (\bx,\tau) j_{\alpha'} (\bx',\tau')
\right\rangle_0$, with $\langle \cdots \rangle_0$ the equilibrium
expectation value, is determined by noting that for the vector
potential ${\bf A} (\bx,\tau) = - c {\bf E} e^{-i \omega
\tau}/\omega$ the above equation
[Eq.~(\ref{eq:expcvaluecurrentPi})] should in the zero-frequency
limit reduce to Ohm's law $\left\langle {\bf j}_{s}^z
\right\rangle_0 = -\hbar (\sigma_\uparrow-\sigma_\downarrow) {\bf
E}/(2 |e|)$. Using this result together with Eqs.~
(\ref{eq:expcvaluecurrentfromeffectiveaction}-\ref{eq:expcvaluecurrentPi}),
we find, after a Wick rotation $\tau \to i t$ to real time, that
\begin{equation}
\label{eq:adiabaticcurrent}
  \langle j_\alpha \rangle = -\frac{\hbar}{2|e|V}\left(
  \sigma_\uparrow\!-\!\sigma_\downarrow \right)
  \frac{\partial}{\partial t} \int d\bx \tilde A_{\alpha'}(\bm{\Omega}(\bx,t)) \nabla_\alpha
  \Omega_{\alpha'} (\bx,t)~,
\end{equation}
with $V$ the volume of the system. We note that the
time-derivative of the Berry phase term is also encountered by
Barnes and Maekawa in discussing the electromotive force in a
ferromagnet \cite{barnes2007}. Such Berry phase terms are known to
occur in adiabatic quantum pumping \cite{zhou2003}.

We now generalize this result to the situation where spin is no
longer conserved, for example due to spin-orbit interactions or
spin-dependent impurity scattering. Linearizing around the
collinear state by means of $\bm{\Omega} \simeq (\delta \Omega_x,
\delta \Omega_y, 1-\delta \Omega_x^2/2-\delta \Omega^2_y/2)$ we
find that the part of the effective action that contains the
electromagnetic vector potential reads \cite{duine2007}
\begin{eqnarray}
\label{eq:actiontransverseflucsgeneral}
 && S^{\rm eff} = \int\!d\tau\!\int\!d\bx\!\int\!d\tau'\!\int\!d\bx'\!\int
  \!d\tau''\!\int\!d\bx'' \left[
  \delta \Omega_a (\bx,\tau) \right.
  \nonumber \\&& \times \left. {\bf K}_{ab}
  (\bx,\bx',\bx'';\tau,\tau',\tau'') \cdot {\bf A} (\bx'',\tau'')
  \delta \Omega_b (\bx',\tau')\right],
\end{eqnarray}
where a summation over transverse indices $a,b\in\{x,y\}$ is
implied. The spin-wave photon interaction vertex
\begin{eqnarray}
\label{eq:intvertex}
  && {\bf K}_{ab}
  (\bx,\bx',\bx'';\tau,\tau',\tau'') =
  \nonumber \\
  &&\frac{\Delta^2}{8\hbar c}
  \langle \phi^\dagger (\bx,\tau) \tau^a \phi(\bx,\tau) \phi^\dagger (\bx',\tau') \tau^b \phi(\bx',\tau') {\bf j} (\bx'',\tau'')
  \rangle_0~,\nonumber \\
\end{eqnarray}
given in terms of the exchange splitting $\Delta$, is also
encountered in a microscopic treatment of spin transfer torques
\cite{duine2007}. The reactive part of this interaction vertex
determines the reactive spin transfer torque and, via
Eqs.~(\ref{eq:expcvaluecurrentfromeffectiveaction}) and
(\ref{eq:actiontransverseflucsgeneral}), reproduces
Eq.~(\ref{eq:adiabaticcurrent}). The zero-frequency
long-wavelength limit of the dissipative part of the spin-wave
photon interaction vertex determines the dissipative spin transfer
torque. (Note that in this approach the definition of the spin
current does not enter in determining the spin transfer torques.)
Although Eq.~(\ref{eq:intvertex}) may be evaluated for a given
microscopic model within some approximation scheme
\cite{duine2007}, we need here only that variation of the action
in Eq.~(\ref{eq:actiontransverseflucsgeneral}) reproduces both the
reactive and dissipative spin torques in
Eq.~(\ref{eq:LLGwithSTTs}). The final result for the electric
current density is then given by
\begin{eqnarray}
\label{eq:finalresultelectriccurrent}  \langle j_\alpha \rangle
&=& -\frac{\hbar}{2|e|V}\left(
  \sigma_\uparrow\!-\!\sigma_\downarrow \right) \left[ \beta  \int d\bx
   \frac{\partial \bm{\Omega} (\bx,t)}{\partial t} \cdot
   \nabla_\alpha \bm{\Omega} (\bx,t)
 \right.\nonumber \\
&&  \left . +  \frac{\partial}{\partial t} \int d\bx \tilde
A_{\alpha'}(\bm{\Omega}(\bx,t)) \nabla_\alpha
  \Omega_{\alpha'} (\bx,t)
  \right]~.
\end{eqnarray}
The above equation is essentially the result of a linear-response
calculation in $\partial \bm{\Omega}/\partial t$, and is the
central result of this paper. We emphasize that the way in which
the transport coefficients $\sigma_\uparrow$ and
$\sigma_\downarrow$ and the $\beta$-parameter enter does not rely
on the specific details of the underlying microscopic model. Note
that the above result reduces to that of Barnes and Maekawa
(Eq.(9) of Ref.~[\onlinecite{barnes2007}]) if we take $\beta=0$.

\section{Field-driven domain wall motion} \label{sec:fddwmotion} To bring out the qualitative physics, we evaluate the
result in Eq.~(\ref{eq:finalresultelectriccurrent}) using a simple
model for field-driven domain wall motion in a magnetic wire of
length $L$. In polar coordinates $\theta$ and $\phi$, defined by
$\bm{\Omega} = (\sin \theta \cos \phi, \sin \theta \sin \phi,\cos
\theta)$, we choose the micromagnetic energy functional
\begin{eqnarray}
\label{eq:energyfunctthetaphinanowire}
  E_{\rm MM} [\theta,\phi] = \rho_s \int d\bx \left\{
    \frac{J}{2} \left[ \left(\nabla \theta \right)^2  + \sin^2 \theta \left(\nabla \phi \right)^2\right]
    \right.
    \nonumber \\
    \left.
    + \frac{K_\perp}{2} \sin^2 \theta \sin^2 \phi - \frac{K_z}{2}
\cos^2 \theta + gB \cos \theta
  \right\}~,
\end{eqnarray}
where $J$ is the spin stiffness, and $K_\perp$ and $K_z$ are
anisotropy constants larger than zero. The external field in the
negative $z$-direction leads to an energy splitting $2 g B>0$. We
solve the equation of motion in Eq.~(\ref{eq:LLGwithSTTs}) within
the variational {\it ansatz} \cite{tatara2004,schryer1974}
\begin{equation}
\label{eq:varansatzdw}
  \theta (\bx,t) = \theta_0 (\bx,t) \equiv 2 \tan^{-1} \left[ e^{- \left( r_{\rm dw} (t) -x
\right)/\lambda} \right]~,
\end{equation}
together with $\phi (\bx,t) = \phi_0 (t)$, that describes a rigid
domain wall with width $\lambda = \sqrt{J/K_z}$ at position
$r_{\rm dw} (t)$. The chirality of the domain wall is determined
by the angle $\phi_0 (t)$ and the magnetization direction is
assumed to depend only on $x$ which is taken in the long direction
of the wire.

The equations of motion for the variational parameters are given
by \cite{tatara2004,duine2006,schryer1974}
\begin{eqnarray}
\label{eq:finaleomvarpars}
  \dot \phi_0 (t) + \alpha_{\rm G} \left( \frac{\dot r_{\rm dw} (t)}{\lambda} \right) =
\frac{gB}{\hbar}~; \nonumber \\
 \left( \frac{\dot r_{\rm dw} (t)}{\lambda} \right) - \alpha_{\rm G}  \dot \phi_0 (t) =
\frac{K_\perp}{2\hbar} \sin 2 \phi_0 (t)~.
\end{eqnarray}
Note that the velocity ${\bf v}_{\rm s}$ is absent from these
equations since we consider the generation of electric current by
a field-driven domain wall. The above equations provide a
description of the field-driven domain wall and, in particular, of
Walker breakdown \cite{schryer1974}. That is, for an external
field smaller than the Walker breakdown field $B_{\rm w} \equiv
\alpha_{\rm G} K_\perp/(2g)$ the domain wall moves with a constant
velocity. For fields $B>B_{\rm w}$ the domain wall undergoes
oscillatory motion, which initially makes the average velocity
smaller.
\begin{figure}
\vspace{-0.5cm} \centerline{\epsfig{figure=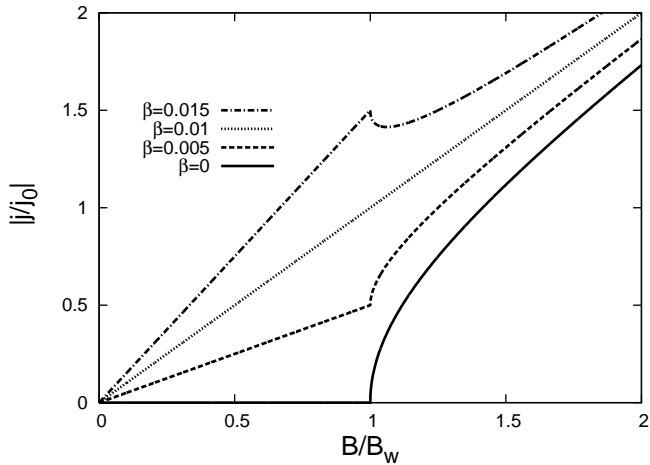}}
 \caption{Current generated by a field-driven domain wall in units of $j_0=2 L/[|e|(\sigma_\uparrow-\sigma_\downarrow)\alpha_{\rm G}
 K_\perp]$, for $\alpha_{\rm G}=0.01$ and various values of $\beta$. The
 result is plotted as a function of magnetic field in units of the
 Walker breakdown field $B_{\rm w}\equiv
\alpha_{\rm G} K_\perp/(2g)$.
  }
 \label{fig:j_b}
\end{figure}

Solving the equations of motion results in
\begin{eqnarray}
\label{eq:soleomvarparsaverage}
 \overline {\dot \phi}_0 &=& \frac{1}{(1+\alpha_{\rm G}^2)} {\rm Re} \left[
 \sqrt{\left( \frac{gB}{\hbar}\right)^2 - \left( \frac{\alpha_{\rm G} K_\perp}{2
 \hbar}\right)^2}\right]~; \nonumber \\
\frac{ \overline {\dot r_{\rm dw}}}{\lambda} &=&
\frac{gB}{\alpha_{\rm G} \hbar} - \frac{\overline {\dot
 \phi}_0}{\alpha_{\rm G}}~,
\end{eqnarray}
where the $\overline{\cdots}$ indicates taking the time-averaged
value. Inserting the variational {\it ansatz} into
Eq.~(\ref{eq:finalresultelectriccurrent}) leads in first instance
to
\begin{equation}
\label{eq:resultcurrentafteransatz}
  \langle j_x \rangle = -\frac{\hbar}{|e| L} \left(\sigma_\uparrow-\sigma_\downarrow
  \right)\left[
  \frac{\beta \dot r_{\rm dw} (t)}{\lambda} + \dot \phi_0 (t)
  \right]~,
\end{equation}
which, using Eq.~(\ref{eq:soleomvarparsaverage}), becomes
\begin{eqnarray}
\label{eq:finalresultwithinapprox} && \overline{\langle j_x
\rangle} = -\frac{\hbar}{|e| L}
\left(\sigma_\uparrow-\sigma_\downarrow \right)
  \left\{ \rule{0mm}{7mm}
   \frac{\beta g B}{\alpha_{\rm G} \hbar} \right. \nonumber \\ && \left. +
   \left( \frac{1- \frac{\beta}{\alpha_{\rm G}}}{1+\alpha_{\rm G}^2} \right) {\rm Re} \left[
 \sqrt{\left( \frac{gB}{\hbar}\right)^2\!\!-\!\left( \frac{\alpha_{\rm G} K_\perp}{2
 \hbar}\right)^2}\right]
  \right\}.
\end{eqnarray}
As shown in Fig.~\ref{fig:j_b}, this result depends strongly on
the ratio $\beta/\alpha_{\rm G}$. In particular, for
$\beta>\alpha_{\rm G}$ a local maximum appears in the current as a
function of magnetic field. Since $\alpha_{\rm G}$ is determined
independently from ferromagnetic resonance experiments,
measurement of the slope of the current for small magnetic fields
enables experimental determination of $\beta$. We note that within
the present approximation the current does not depend on the
domain wall width $\lambda$. Furthermore, in the limit of zero
Gilbert damping and $\beta$, the dissipationless limit, we have
that the current density is equal to $\overline{\langle j_x
\rangle} = \left(\sigma_\downarrow-\sigma_\uparrow \right) gB/(|e|
L)$. This is the result of Barnes and Maekawa \cite{barnes2007}
that corresponds to the situation that $\alpha_{\rm G} = \beta
=0$, as discussed in the Introduction. We point out that, within
our approximation for the description of domain-wall motion,
putting $\beta=\alpha_{\rm G}$ in
Eq.~(\ref{eq:finalresultwithinapprox}) gives the same result as
using
Eqs.~(\ref{eq:finaleomvarpars})~and~(\ref{eq:resultcurrentafteransatz})
with $\alpha_{\rm G}=\beta=0$. That the situation discussed by
Barnes and Maekawa \cite{barnes2007} is indeed that of
$\alpha_{\rm G} = \beta =0$ is seen by comparing their result
[Eqs.~(8)~and~(9) of Ref.~[\onlinecite{barnes2007}], and the
paragraph following Eq.~(9)] with our results in
Eqs.~(\ref{eq:finalresultelectriccurrent})~and~(\ref{eq:finaleomvarpars}).

\section{Discussion and conclusions} \label{sec:disc} Our result in
Eq.~(\ref{eq:finalresultwithinapprox}) is a simple expression for
the pumped current as a function of magnetic field for a
field-driven domain wall. A possible disadvantage in using
Eq.~(\ref{eq:finalresultwithinapprox}), however, is that in
deriving this result we assumed a specific model to describe the
motion of the domain wall. This model does in first instance not
include extrinsic pinning and nonzero temperature. Both extrinsic
pinning \cite{tatara2004} and nonzero temperature \cite{duine2006}
can be included in the rigid-domain wall description. However, it
is in some circumstances perhaps more convenient to directly use
the result in Eq.~(\ref{eq:resultcurrentafteransatz}) together
with the experimental determination of $\dot r_{\rm dw} (t)$.
Since the only way in which the parameter $\beta$ enters this
equation is as a prefactor of $\dot r_{\rm dw} (t)$, this should
be sufficient to determine its value from experiment. We note,
however, that the precision with which the ratio
$\beta/\alpha_{\rm G}$ can be determined depends on how accurately
the magnetization dynamics, and, in particular, the motion of the
domain wall, is imaged experimentally. With respect to this, we
note that the various curves in Fig.~\ref{fig:j_b} are
qualitatively different for different values of $\beta/\alpha_{\rm
G}$. In particular, the results for $\beta/\alpha_{\rm G}>1$ and
$\beta/\alpha_{\rm G}<1$ differ substantially, and could most
likely be experimentally distinguished. In view of this
discussion, future research will in part be directed towards
evaluating Eq.~(\ref{eq:finalresultelectriccurrent}) for more
complicated models of field-driven domain-wall motion, which will
benefit the experimental determination of $\beta/\alpha_{\rm G}$.

A typical current density is estimated as follows. For the
experiments of Beach {\it et al.} \cite{beach2005} we have that
$L\sim 20$ $\mu$m, and $\lambda \sim 20$ nm. The domain velocities
measured in this experiment are $\dot r_{\rm dw}\sim40-100$ m/s.
Taking as a typical conductivity $\sigma_\uparrow \sim 10^6$
$\Omega^{-1}$m$^{-1}$ we find, using equation
Eq.~(\ref{eq:resultcurrentafteransatz}) with $\beta \sim 0.01$,
typical electric current densities of the order of $\langle j_x
\rangle \sim 10^{3}-10^4$ A m$^{-2}$. This result depends somewhat
on the polarization of the electric current in the ferromagnetic
metal, which we have taken equal to $50\%-100\%$ in this rough
estimate. Although much smaller than typical current densities
required to move the domain wall via spin transfer torques,
electrical current densities of this order appear to be detectable
experimentally.

In conclusion, we have presented a theory of spin pumping without
spin conservation, and, in particular, proposed a way to gain
experimental access to the parameter $\beta/\alpha_{\rm G}$ that
is of great importance for the physics of current-driven domain
wall motion. We note that the mechanism for current generation
discussed in this paper is quite distinct from the generation of
eddy currents by a moving magnetic domain \cite{colaiori2007}. In
addition to improving upon the model used for describing
domain-wall motion, we intend to investigate in future work
whether the damping terms in Eq.~(\ref{eq:LLGwithSTTs}), or
possible higher-order terms in frequency and momentum
\cite{footnote}, have a natural interpretation in terms of spin
pumping, similar to the spin-pumping-enhanced Gilbert damping in
single-domain ferromagnets \cite{tserkovnyak2002}.

It is a great pleasure to thank Gerrit Bauer, Maxim Mostovoy, and
Henk Stoof for useful comments and discussions.


\begin{thebibliography}{99}
\bibitem{footnote1} We note that the dissipative spin transfer torque is
sometimes referred to as ``non-adiabatic". Both the reactive and
dissipative spin transfer torques in Eq.~(\ref{eq:LLGwithSTTs})
are adiabatic in the sense that they arise to lowest order in a
gradient expansion of the magnetization direction.
\end{thebibliography}
\end{document}